\documentclass[preprint,11pt]{elsarticle}
\usepackage{amsmath,amsthm,amsfonts,amssymb}
\usepackage[mathcal]{eucal}
\usepackage{mathrsfs}
\usepackage{graphicx}
 \usepackage{graphics}
\usepackage{dcolumn}
\usepackage{latexsym}
\usepackage{epsfig}
\usepackage{bm}
\usepackage[all]{xy}
\usepackage[T1]{fontenc}

\begin{document}

\begin{frontmatter}

\title{Thermodynamical properties of graphene in noncommutative phase-space}

\author[UFC,Victor]{Victor Santos}
\ead{vsantos@gravity.psu.edu}

\author[UFC]{R. V. Maluf}
\ead{r.v.maluf@fisica.ufc.br}

\author[UFC]{C. A. S. Almeida}
\ead{carlos@fisica.ufc.br}

\address[UFC]{Departamento de F\'{i}sica - Universidade Federal do Cear\'{a} (UFC) - C.P. 6030, 60455-760
Fortaleza-Cear\'{a}-Brazil}

\address[Victor]{Institute for Gravitation and the Cosmos \& Physics Department,
Pennsylvania State University, University Park, PA 16801, U.S.A.}

\begin{abstract}
We investigated the thermodynamic properties of graphene in a noncommutative phase-space in the presence of a constant magnetic field. In particular, we determined the behaviour of the main thermodynamical functions: the Helmholtz free energy, the mean energy, the entropy and the specific heat. The high temperature limit is worked out and the thermodynamic quantities, such as
mean energy and specific heat, exhibit the same features as the
commutative case. Possible connections with the results already
established in the literature are discussed briefly.
\end{abstract}






\end{frontmatter}


\section{Introduction\label{sec:introduction}}

Carbon, in its allotropic forms like graphite and diamond, takes up a prominent place in different branches of science. In particular, graphite can be thought as composed by stacking one-atom thick layers of carbon, called \emph{graphene}. The physics of graphene has attracted attention from theoretical scientific community since experimental observations revealed the existence of electrical charge carriers that behave as massless Dirac quasi-particles  \cite{Novoselov2004,Novoselov2005,CastroNeto2009,Peres2010}.
The reason for this is due to the unusual molecular structure of graphene.
The Carbon atoms are arranged in a hexagonal lattice, similar to a honeycomb structure \cite{Fefferman2012}.
It was observed that the low-energy electronic excitations at the
corners of graphene Brillouin zone can be described by a $2+1$
Dirac fermions with linear dispersion relation (massless) \cite{CastroNeto2009,Peres2010}.
This effect offers the prospect of testing several aspects of relativistic
phenomena, which usually requires large energy, in experiments of
the condensed matter physics such as chiral tunneling and Klein paradox
\cite{Katsnelson2006,Katsnelson2007}.

On the other side, the study of quantum systems in a noncommutative (NC) space has been the subject of much interest in last years,  assuming that  noncommutativity may be, in fact, a result of quantum gravity effects \cite{NoncomutativeRef}.
In these studies, some attention has been given to the models of noncommutative quantum mechanics (NCQM) \cite{NCQM1}. The interest in this approach lies on the fact that NCQM is a fruitful theoretical laboratory where we can get some insight on the consequences of noncommutativity in field theory by using standard calculation techniques of quantum mechanics. In this context, several types of noncommutativity have been considered \cite{NCMQ2} and one case of particular importance is the so called noncommutative phase-space. This specific formulation is necessary to implement the Bose-Einstein statistics in the context of NCQM \cite{Nair2001, Zhang2004}.

The NC phase-space is based on the assumption that the spatial coordinates
$\hat{x}_{i}$ and the conjugate momenta $\hat{p}_{i}$ are operators
satisfy a deformed Heisenberg algebra, which in its simplest form can
be described by the commutation relations:
\begin{equation}
[\hat{x}_{i},\hat{x}_{j}]=i\theta_{ij},\ \ \ [\hat{p}_{i},\hat{p}_{j}]=i\eta_{ij},\ \ \ [\hat{x}_{i},\hat{p}_{j}]=i\hbar\left(\delta_{ij}+\frac{\theta_{ik}\eta_{jk}}{4\hbar^{2}}\right),\ \ \ \mbox{with}\ i,j,k=1,\dots\mbox{d},\label{eq:1}
\end{equation}
where the deformation parameters $\theta_{ij}=\theta\epsilon_{ij}$ and $\eta_{ij}=\eta\epsilon_{ij}$ are
real and antisymmetric constants matrices. These commutation relations can be explicitly implemented by means of coordinate transformations \cite{Bertolami2011}:
\begin{equation}
\hat{x}_{i}=x_{i}-\frac{\theta_{ij}}{2\hslash}p_{j},\ \ \ \hat{p}_{i}=p_{i}+\frac{\eta_{ij}}{2\hslash}x_{j},\label{Noncoordinate}
\end{equation}
where $x_{i}$ and $p_{i}$ are commutative variables that satisfy
ordinary Heisenberg commutation relations,
\begin{equation}
[x_{i},x_{j}]=0,\ \ \ [p_{i},p_{j}]=0,\ \ \ [x_{i},p_{j}]=i\hslash\delta_{ij}.
\end{equation}

Recently, graphene in the framework of NCQM was studied by Bastos et al. \cite{Bastos2013}, where the authors determined the Hamiltonian and the associated energy spectrum. It was shown that the electron states close to the Dirac points ($K$ and $K^{'}$ points at the corners of graphene Brillouin zone) in a NC phase-space, subject to an external constant magnetic field, can be described by a massless $2D$ Dirac equation with only momenta noncommutativity. Otherwise,
we would have a gauge symmetry breaking, which it is not observed in
the graphene lattice \cite{Bertolami2011}.

These results, in association with suitable experimental data, may be used
to investigate the role of noncommutativity in the graphene system and improve bounds on the magnitude of the corresponding noncommutative parameters. For instance, the issue concerning the thermodynamics properties of graphene modified
by this kind of theory has not been addressed. Thus, using an approach similar to the cases of Dirac and Kemmer oscillators studied in Refs. \cite{Pacheco2003,Boumali2007}, we propose to evaluate the main thermodynamic functions that describe the thermal behaviour of this system in both cases; commutative and noncommutative. One such study with a focus on graphene is particularly interesting because this material has numerous applications for thermal industry, and it may be important in the understanding of heat conduction in low dimensions \cite{Balandin2011,Pop2012,Alofi2013}.

This work is outlined as follows. In Sec. \ref{Energy}, we summarize the key results of Ref. \cite{Bastos2013} which we will use in the sequel. In Sec. \ref{Calc}, the solution for the energy levels is utilized to calculate the partition function, and then all thermodynamic quantities that describe the thermal physics of NC graphene. The methodology used closely follows that developed in Refs. \cite{Pacheco2003,Boumali2007}. Finally, in sec. \ref{Conclusion}, we present the conclusion and final remarks.


\section{\label{Energy}Graphene in a noncommutative phase-space}

Before studying the thermal properties of graphene, let us first recall
the fundamentals on the graphene physics in a NCQM approach. Here, we follow
the steps described in Ref. \cite{Bastos2013}. The
basic equation in the theory is the Dirac equation for a free massless
particle:
\begin{equation}
i\hslash\frac{\partial\psi}{\partial t}=H_{D}\psi,
\end{equation}
where $H_{D}$ represents two copies of the massless Dirac-like Hamiltonian that describes the behaviour of electrons around each Dirac points $K$ and $K'$, at the corners of Brillouin zone.

From this way, we can explicitly write:
\begin{equation}
H_{D}=\begin{pmatrix}H_{K} & 0\\
0 & H_{K^{'}}
\end{pmatrix}=v_{F}\begin{pmatrix}\boldsymbol{\sigma}\cdot\boldsymbol{p} & 0\\
0 & \boldsymbol{\sigma}^{*}\cdot\boldsymbol{p}
\end{pmatrix},\mbox{ and }\psi=\begin{pmatrix}\psi_{K}\exp(-\frac{i}{\hslash}E_{K}t)\\
\psi_{K'}\exp(-\frac{i}{\hslash}E_{K'}t)
\end{pmatrix},\label{eq:Hamiltoniandef}
\end{equation}
such that $v_{F}\backsimeq10^{6}\text{m/s}$ is the Fermi velocity, $\boldsymbol{p}=-i\hslash(\partial_{x},\partial_{y},0)$
is the two-dimensional momentum operator, $\boldsymbol{\sigma}=(\sigma_{x},\sigma_{y},\sigma_{z})$
corresponds to the Pauli spin matrices and $\psi_{K(K')}$ are two-component
wavefunction close to the $K$ $(K')$ point. Besides, $E_{K(K')}=\pm\hslash v_{F}\left|\boldsymbol{k}\right|$ represents the eigenvalues associated with the positive/negative energy band.

Considering an external homogeneous magnetic field, ${\bf B}=B_0\hat{z}$,
we must make the usual minimal substitution ${\bf p}\rightarrow{\bf p}-e{\bf A}$
in the free Hamiltonian \eqref{eq:Hamiltoniandef}, such that
\begin{equation}
H_{K}=v_{F}\sigma_{i}(p_{i}+\frac{eB_0}{2}\epsilon_{ij}x_{j}),\ \ \ \ i,j=1,2,
\label{HamGraf}\end{equation}
where the vector potential is written in the symmetric gauge ${\bf A}=\frac{B_0}{2}(-y,x,0).$ A similar equation can also be obtained for the Dirac point $K'$.

We can diagonalize the Hamiltonian \eqref{HamGraf} with an appropriate set of annihilation and creation operators, and obtain the following expression for the energy eigenvalues \cite{Bastos2013}
\begin{equation}
E_{K}=\pm\frac{\sqrt{2}\hbar v_{F}}{l_{B}}\sqrt{n},\quad n=0,1,2,\dots,\label{eq:Energyspectrum}
\end{equation}
where $l_{B}=\sqrt{\hbar/eB_0}$ is called magnetic length.

On NC phase-space the Dirac Hamiltonian of graphene can be achieved through the shift given in Eq. \eqref{Noncoordinate}.  However, in order to preserve the gauge invariance on the graphene sheet, it should be required that the noncommutativity is nonzero only in momentum coordinates \cite{Bertolami2011}. Thus, the Eq. \eqref{HamGraf} becomes
\begin{equation}
H_{K}^{\mbox{NC}}=v_{F}\sigma_{i}(p_{i}+\frac{eB_0}{2}\mu\epsilon_{ij}x_{j}),\ \ \ \ i,j=1,2,
\end{equation}
where
\begin{equation}
\mu=\left(1+\frac{\eta}{eB_{0}\hslash}\right).
\end{equation}

Finally, it is possible to show that the energy spectrum takes the following form
\begin{equation}
E_{K}=\pm\frac{\sqrt{2}\hbar v_{F}}{l_{B}}\sqrt{(1+\bar{\eta})n},\quad n=0,1,2,\dots,\label{eq:EnergyspectrumNC}
\end{equation}
with $\bar{\eta}$ being a dimensionless constant related to the noncommutative parameter $\eta$ by means of $\bar{\eta}=(l_{B}^{2}/\hbar^{2})\eta$.


\section{\label{Calc}Thermal properties of graphene in a noncommutative phase-space}

\subsection{Theoretical framework}

In working out the thermal properties of graphene system, let us begin by defining the fundamental object in statistical mechanics, the canonical partition function $Z$. Given the energy spectrum of graphene, we can define the partition function by a sum over all states $s$ of the system, via
\begin{equation}
Z=\sum_{s}\exp(-\beta E_{s}),\quad\beta=\frac{1}{k_{B}T},
\end{equation}
where $k_{B}$ is the Boltzmann constant and $T$ is the equilibrium
temperature.

The entire thermodynamics of NC graphene can be derived from the partition function $Z$ and the Eq. \eqref{eq:EnergyspectrumNC}. The most important thermodynamic functions for our analysis are the Helmholtz free energy $F$, the mean energy $U$, the entropy $S$ and the specific heat $C$, defined by the following expressions:\begin{align}
F & =-\frac{1}{\beta}\ln Z,\\
U & =-\frac{\partial}{\partial\beta}\ln Z,\\
S & =k_{B}\beta^{2}\frac{\partial F}{\partial\beta},\\
C & =-k_{B}\beta^{2}\frac{\partial U}{\partial\beta}.
\end{align}

We are now in condition to perform the numerical calculations of the partition function and all other thermodynamics quantities. To this end, it is convenient to introduce the dimensionless variables,\begin{equation}
\bar{\beta}=\frac{1}{\tau},\quad \tau=\frac{l_{B}}{\sqrt{2}\hbar v_{F}}\frac{1}{\beta},
\end{equation}
and rewrite all the previous functions in terms of these new quantities. {In this case, we get $F\rightarrow\bar{F}=\frac{l_{B}}{\sqrt{2}\hbar v_{F}}F=-\frac{1}{\bar{\beta}}\log Z$, $U\rightarrow\bar{U}=\frac{l_{B}}{\sqrt{2}\hbar v_{F}}U=-\frac{\partial}{\partial\bar{\beta}}\log Z$, $S\rightarrow\bar{S}=\frac{S}{k_{B}}=\bar{\beta}^{2}\frac{\partial\bar{F}}{\partial\bar{\beta}}$ and $C\rightarrow\bar{C}=\frac{C}{k_{B}}=-\bar{\beta}^{2}\frac{\partial\bar{U}}{\partial\bar{\beta}}$, which are now all dimensionless functions. The remainder of the calculation procedure follows in a similar fashion to that described in Refs. \cite{Pacheco2003,Boumali2007}.

Using the Eq. \eqref{eq:EnergyspectrumNC}, the partition function reads
\begin{equation}
Z=1+\sum_{n=0}^{\infty}e^{-\bar{\beta}\sqrt{an+a}},\label{Sum1}
\end{equation}
with $a=1+\bar{\eta}$. Let us note that the definition for the parameter $a$ encompasses the noncommutative as well as the commutative case $(\bar{\eta}=0)$. The integral test ensures the convergence of the series, since the integral\begin{equation}
\int_{0}^{\infty}\textup{d}x\,e^{-\bar{\beta}\sqrt{ax+a}}=\frac{2}{a\bar{\beta}^{2}}e^{-\bar{\beta}\sqrt{a}}\Big(1+\bar{\beta}\sqrt{a}\Big)
,\end{equation}
is finite. To evaluate the sum in \eqref{Sum1} and determine the high temperature limit, we will employ the Euler-Mclaurin summation formula given by
\begin{eqnarray}
\sum_{n=a}^{b}f(n) & = &  \int_{a}^{b}\textup{d}t\,f(t)+\frac{1}{2}\big[f(b)+f(a)\big]+\sum_{i=2}^{k}\frac{b_{i}}{i!}\big[f^{(i-1)}(b)-f^{(i-1)}(a)\big]\nonumber\\
 & - &\int_{a}^{b}\textup{d}t\,\frac{B_{k}\big(\{1-t\}\big)}{k!}f^{(k)}(t),\label{eq:summationformula}
\end{eqnarray}
where $a$ and $b$ are arbitrary real numbers with difference $b-a$ being a positive integer number. $B_{k}$ and $b_{n}$ are Bernoulli polynomials
and Bernoulli numbers, respectively, and $k$ is any positive integer. The symbol $\{x\}$ for a real number $x$ denotes the fractional part of $x$ \cite{Abramowitz}.

According to the relations \eqref{Sum1} and \eqref{eq:summationformula}, and considering $f(n)=\exp\left[-\bar{\beta}\sqrt{an+a}\right]$, the partition function $Z$ can explicitly be written as
\begin{equation}
Z=1+\frac{1}{2}e^{-\bar{\beta}\sqrt{a}}+\frac{2}{a\bar{\beta}^{2}}e^{-\bar{\beta}\sqrt{a}}\Big(1+\bar{\beta}\sqrt{a}\Big)-\sum_{i=2}^{k}\frac{b_{i}}{i!}f^{(i-1)}(0)+R_{k},\label{Z2}
\end{equation}
with
\begin{equation}
R_{k}=\int_{0}^{\infty}\textup{d}t\,\frac{B_{k}\big(\{1-t\}\big)}{k!}f^{(k)}(t),\label{eq:remainder_term}
\end{equation}
being the remainder term. The accuracy of approximation in \eqref{Z2} depends on the asymptotic behaviour of the remainder term $R_k$ while $k$ goes to infinity. In order to get a glimpse of the convergence speed,  we display the behaviour of $R_k$ in Figures \ref{fig:Remainder1} and \ref{fig:Remainder2}, where the number of terms is $k\leq25$. Let us note that the remainder term quickly decreases, becoming of the order of $10^{-5}$ after the summation of 10 terms of the series, and hence, for our purpose it is sufficient to take the summation of $k=50$ in Euler-Maclaurin expansion for the partition function.
\begin{figure}[!htb] 
       \begin{minipage}[b]{0.48 \linewidth}
           \includegraphics[width=\linewidth]{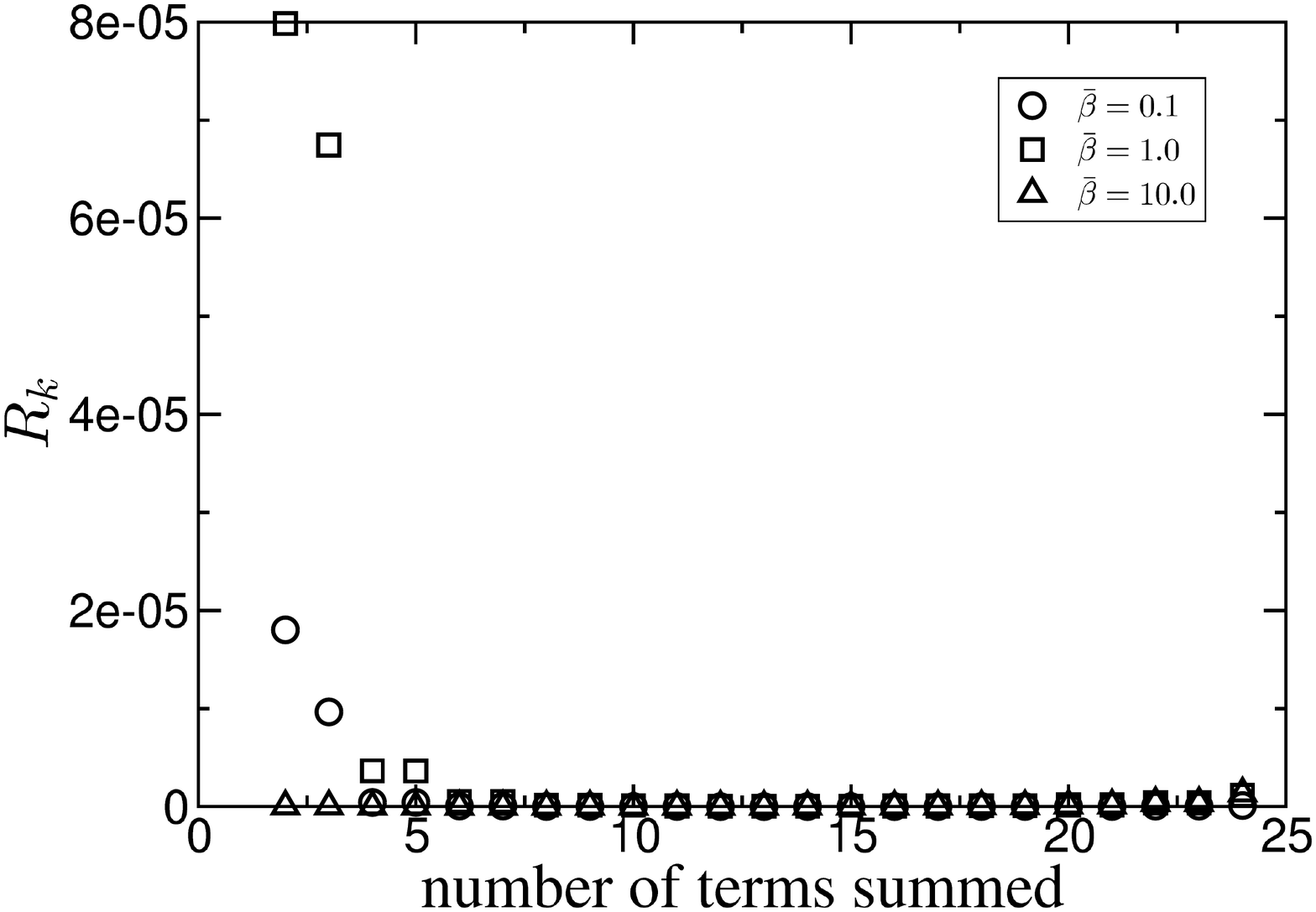}\\
           \caption{Plots of the remainder term $R_{k}$ for the commutative case $\bar{\eta}=0$.}
          \label{fig:Remainder1}
       \end{minipage}\hfill
       \begin{minipage}[b]{0.48 \linewidth}
           \includegraphics[width=\linewidth]{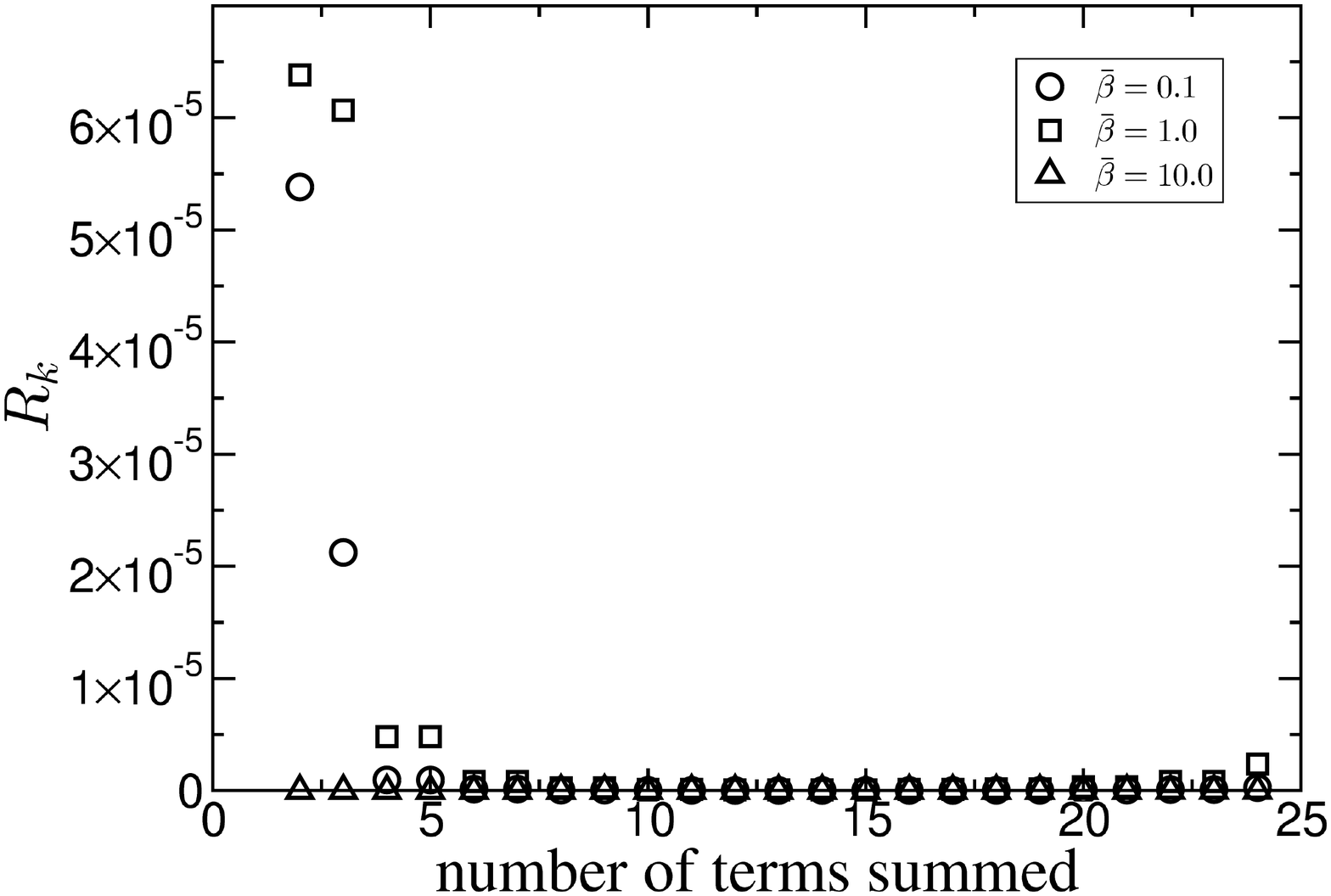}\\
           \caption{Plots of the remainder term $R_{k}$ for the NC case with $\bar{\eta}=0.1$.}
           \label{fig:Remainder2}
       \end{minipage}
   \end{figure}

Before discussing the main results, we consider what happens in the higher temperatures limit to the case of the mean energy $U$ and the specific heat $C$. In this regime,  the dimensionless parameter $\bar{\beta}$ is very small and we can approximate the partition function by
\begin{equation}
Z\simeq\frac{2}{a\bar{\beta}^{2}},\label{eq:partition_function_asymp}
\end{equation}which leads to the following asymptotic limits
\begin{eqnarray}
\bar{U}& \simeq & \frac{2}{\bar{\beta}},\label{eq:mean_energy_asymp}\\
\bar{C}&\simeq & 2,\label{eq:heat_capacity_asymp}
\end{eqnarray}where the bar denotes dimensionless quantities. We note that the dependence on the noncommutative parameter $\bar{\eta}$ is removed. Besides, these limits follow the Dulong-Petit law for an ultra-relativistic ideal gas. In such case, the average energy and specific heat are twice that of the non-relativistic limit. It is worth mentioning that similar results were observed for the Dirac and Kemmer oscillators in a thermal bath \cite{Pacheco2003,Boumali2007}.

\subsection{Results and discussions}

We are now ready to present our calculations for the thermodynamic properties of graphene, in a NC phase-space. All quantities were computed as a function of the dimensionless temperature $\tau$, defined by $\tau=(l_{B}/\sqrt{2}\hbar v_{F})k_{B}T$. The only free parameter that remains to be fixed is $\bar{\eta}$, and it is linked to the momentum noncommutative parameter $\eta$ by the relation $\bar{\eta}=(l_{B}^{2}/\hbar^{2})\eta$. Several bounds for the magnitude of this $\eta$-parameter have been determined in the literature. For example, in Ref. \cite{Bertolami2011} a stringent bound on $\eta$, namely, $\sqrt{\eta}\lesssim 2.26\times10^{-6} \mbox{eV/c}$ was obtained, with the analysis of the hyperfine structure of hydrogen. Concerning the graphene system in the presence of a magnetic field, infrared spectroscopy experiments can be used to establish an upper bound of $\bar{\eta}<0.069$, which implies that the magnitude of the parameter $\eta$ not be larger than $\sqrt{\eta}<8.6 \mbox{eV/c}$ \cite{Bastos2013}. This result is less restrictive than the early bound, however, it is more consistent with typical energy scales and experimental setups of graphene. Thereby, in the present approach, we will assume that the noncommutative effects are relevant at the interval $0<\bar{\eta}<0.1$. In order to illustrate the behavior of the thermodynamic functions with respect to the parameter $\bar{\eta}$, let us consider the four different values $\bar{\eta}=0.00$ (commutative case), $0.01$, $0.04$ and $0.07$.

As a general result, we observed that the noncommutativity effects are very small and in the limit $\bar{\eta}\rightarrow0$ all curves agree with the commutative case. The Helmholtz free energy $\bar{F}$ for the commutative (solid line) and noncommutative (dashed lines) systems are shown in Figure \ref{fig:FreeEnergy}. In any case it decreases with temperature, as expected. One observes that the free energies coincide at low temperatures, but occurs a \textbf{tiny} separation that increases with temperature and with the magnitude of the noncommutative parameter $\bar{\eta}$. From Figure \ref{fig:AverageEnergy}, we can observe the behaviour of the mean energy with temperature. The plots of the mean energy, for different values of $\bar{\eta}$, do not differ significantly, and at the  asymptotic limit,  all the curves behave as linear functions of  temperature. This is an expected result  in view of our previous discussion.

Figure \ref{fig:Entropy} provides the entropy curves of graphene for some values of $\bar{\eta}$. As in the case of the free energy, the behaviour of the entropy for commutative and noncommutative cases is similar at low temperatures.  Note that the separation of the curves increases when the parameter $\bar{\eta}$ grows. Here, the effect of noncommutativity is to deviate to a lower value the entropy, in contrast to the free energy. Indeed, note that the curves with $\bar{\eta}\neq 0$ are located below the solid curve which represents the commutative case. This effect is expected because it is known that the noncommutativity leads to a decrease of the degeneracy in physical systems, reflecting a reduction of the entropy in a NC phase-space \cite{Mitra2002,Saadat2013}.

Finally, let us analyze the behaviour of the specific heat as shown in Figure \ref{fig:HeatCapacity}. Our calculations reveal that the noncommutative profiles change in the intermediate temperature range (about $\tau = 0.95$, approximately). We note that for low $\tau$ values the dashed lines are located  below the solid line while the opposite occurs for large $\tau$ values (see insertion with broadened plots in Fig. \ref{fig:HeatCapacity}). In particular, at high temperature regime all curves coincide with the limit value obtained in Eq. \eqref{eq:heat_capacity_asymp}.


\section{\label{Conclusion}Conclusion}

In the present work, we have studied the thermodynamic properties
of graphene in a NC phase-space with the presence of an
external constant magnetic field. Here, starting from the well-known Dirac
Hamiltonian and its noncommutative extension presented in Ref. \cite{Bastos2013}, we used the solutions of the energy spectrum
in the numerical evaluation of the canonical partition function.

In the sequel, we have determined the main thermodynamic functions
in both cases, commutative and noncommutative. The results are depicted in Figures \ref{fig:FreeEnergy}, \ref{fig:AverageEnergy},
\ref{fig:Entropy} and \ref{fig:HeatCapacity}. The overall behaviour of the thermodynamic functions indicates that the noncommutativity produces a small deviation around the usual commutative profile, and this deviation is positive for the free energy and negative for the entropy. Moreover, the curves for the mean energy and the specific heat are in agreement with
the asymptotic values as given in Eqs. \eqref{eq:partition_function_asymp}
and \eqref{eq:heat_capacity_asymp}, respectively.

Finally, we expect in the near future, once obtained experimental measurements of high accuracy involving the thermodynamical properties of graphene, our
results may be used as a good tool to study these properties. Also, we intend to investigate the possibility of new features of graphene can be described by NC models.


             \begin{figure}[!htb] 
                  \includegraphics[scale=0.6]{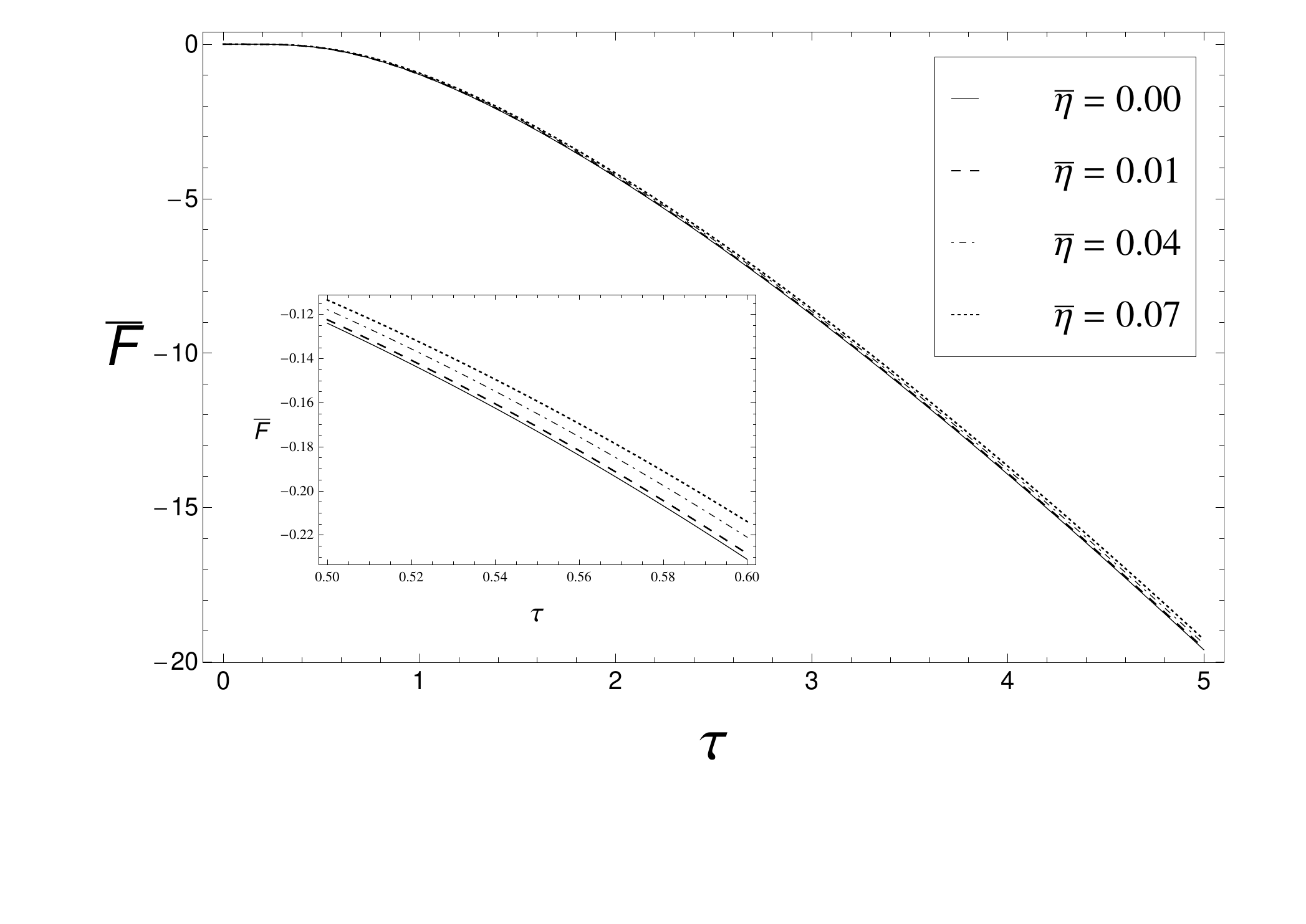}\\
           \caption{The free energy for the
Dirac particle as a function of the dimensionless temperature $\tau$,
for different values of the noncommutativity parameter $\bar{\eta}$.\label{fig:FreeEnergy}}
             \end{figure}
             \begin{figure}[!htb] 
                  \includegraphics[width=\linewidth]{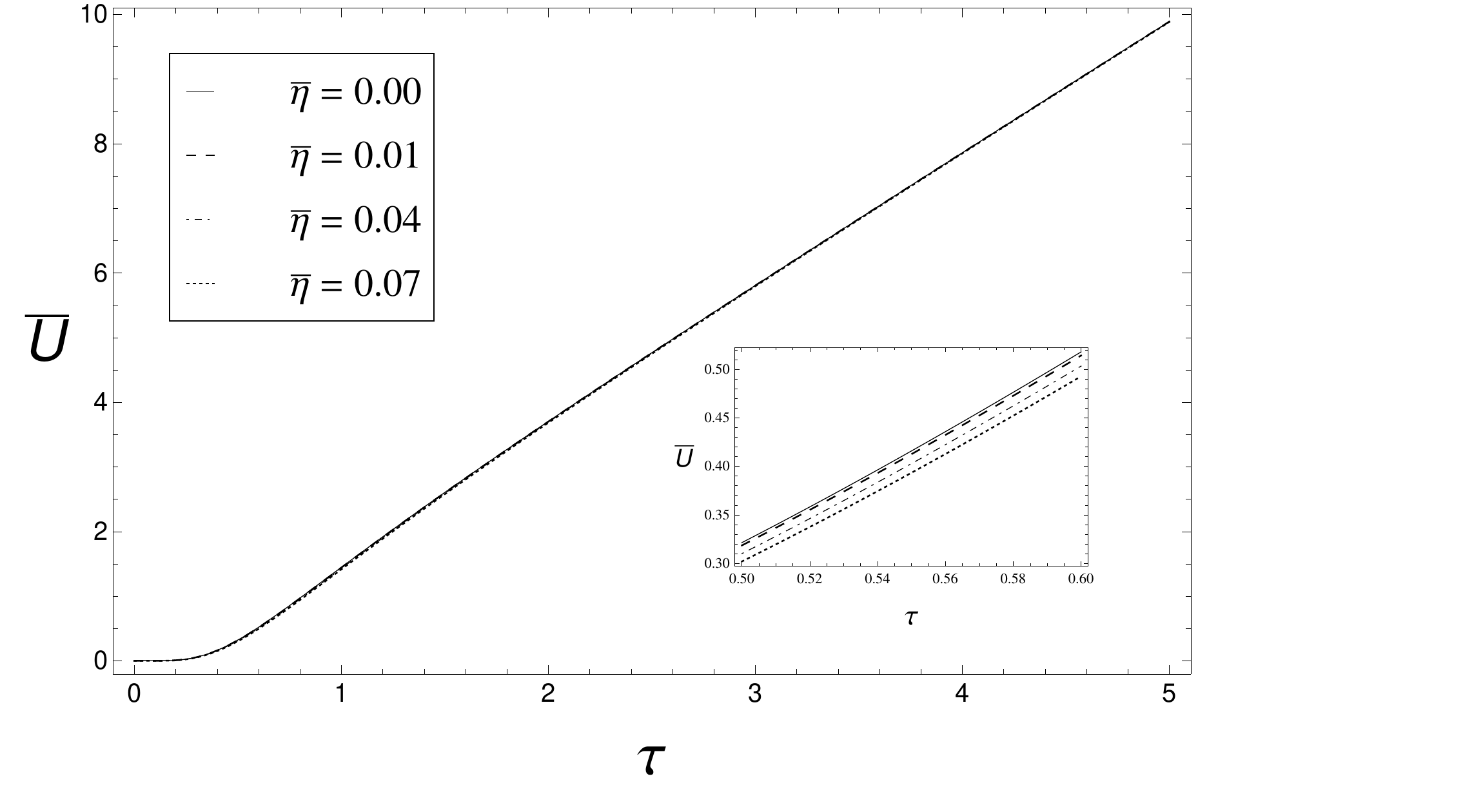}\\
           \caption{The mean energy as
a function of the dimensionless temperature $\tau$, for different
values of the noncommutativity parameter $\bar{\eta}$.\label{fig:AverageEnergy}}
             \end{figure}
             \begin{figure}[!htb] 
                  \includegraphics[width=\linewidth]{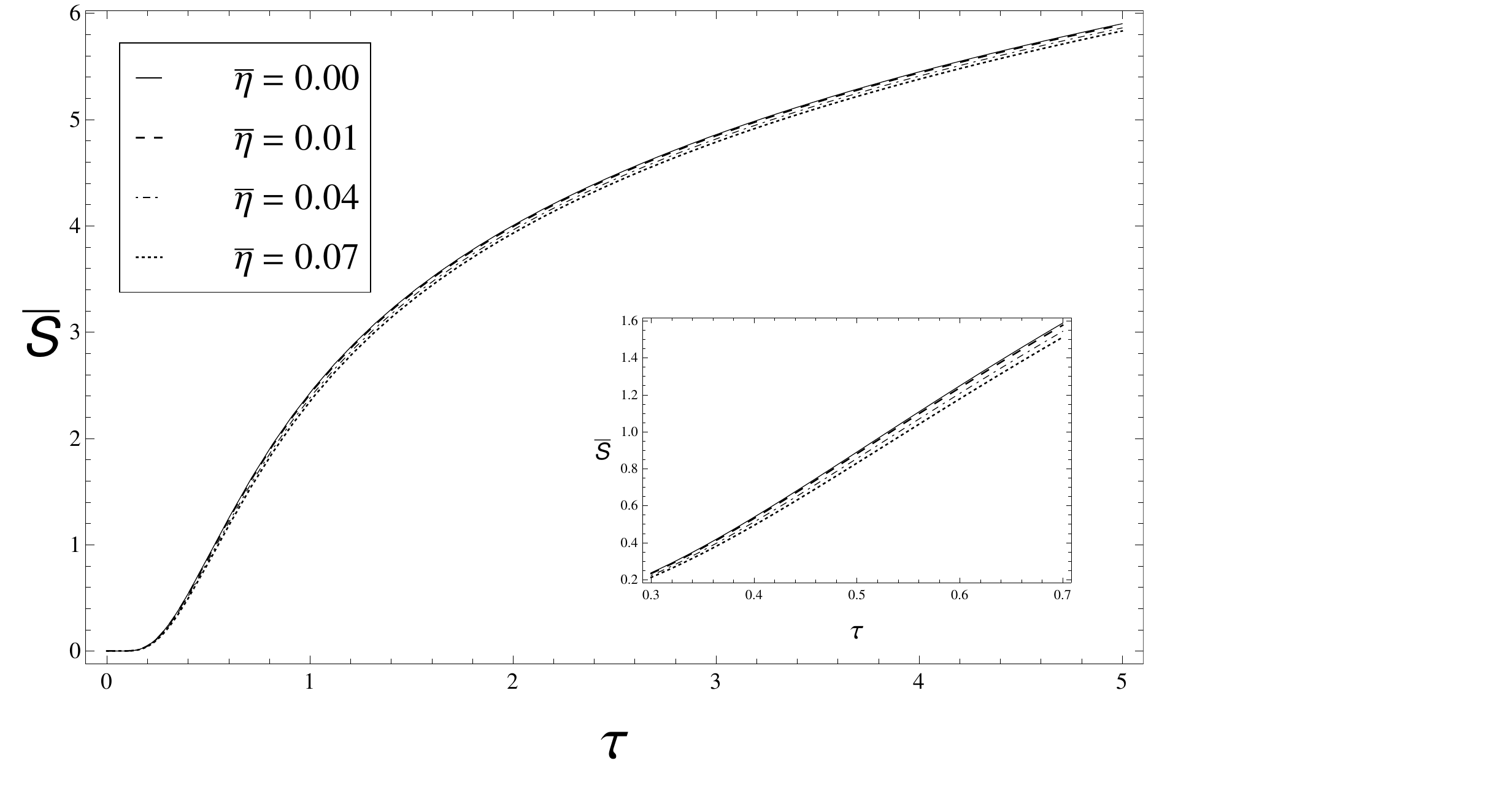}\\
           \caption{The entropy as a function of the dimensionless
temperature $\tau$, for different values of the noncommutativity
parameter $\bar{\eta}$.\label{fig:Entropy}}
             \end{figure}
\begin{figure}[hhtb] 
                  \includegraphics[width=\linewidth]{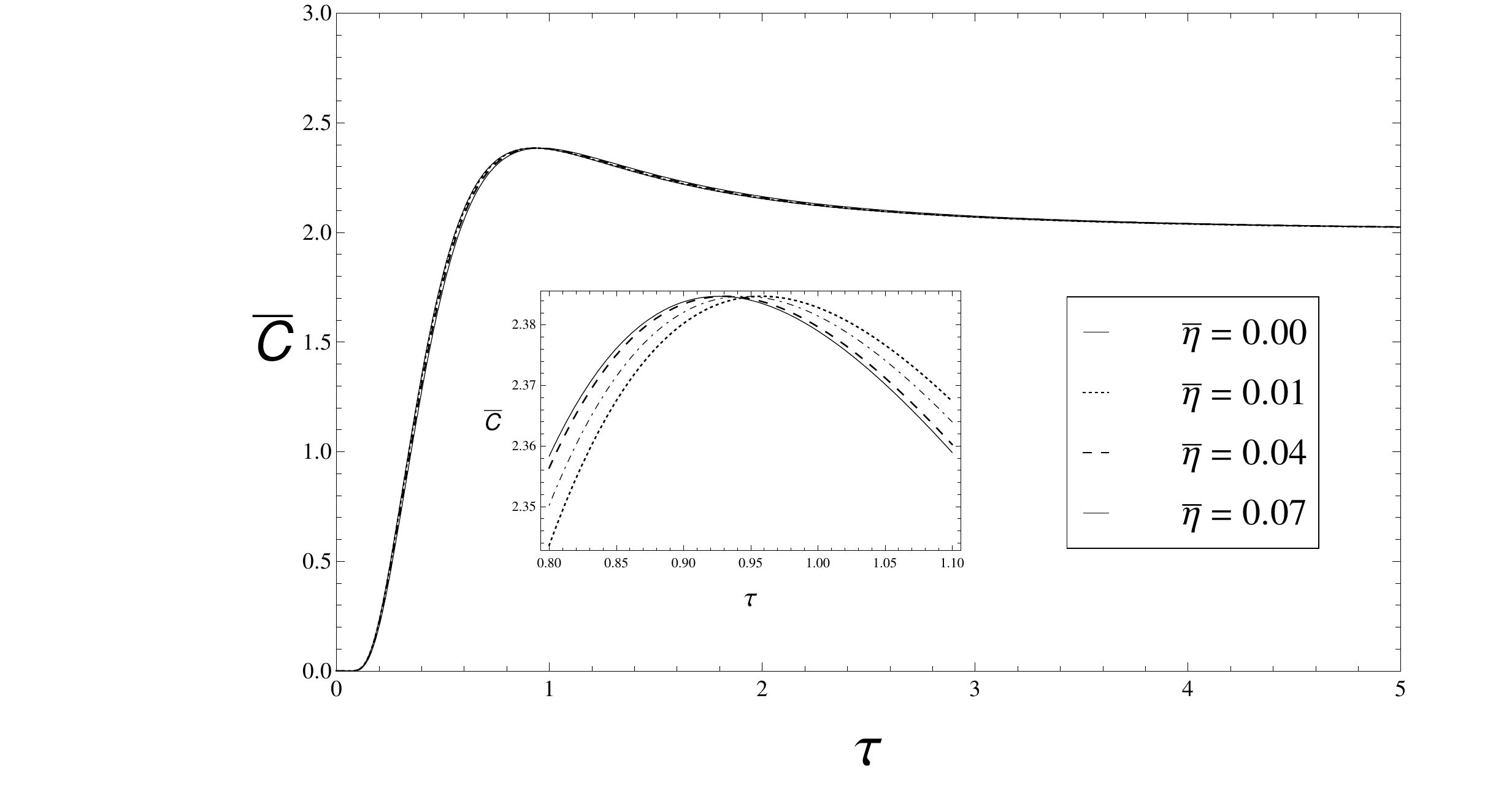}\\
           \caption{The specific heat as a function of the dimensionless temperature $\tau$, for different values of the
noncommutativity parameter $\bar{\eta}$.\label{fig:HeatCapacity}}
             \end{figure}


\section*{Acknowledgments}
The authors thank the Coordena\c{c}\~ao de Aperfei\c{c}oamento de Pessoal de
N\'{i}vel Superior (CAPES), and the Conselho Nacional de Desenvolvimento
Cient\'{i}fico e Tecnol\'ogico (CNPq) for financial support.


\end{document}